\documentclass[fleqn,twoside]{article}

\usepackage{espcrc2}                                 % Elsevier style
\usepackage{graphicx}
\usepackage{amsmath,amsfonts}
\usepackage[numbers,sort&compress]{natbib}           % for [1,2,3] --> [1-3] 
\graphicspath{{./Figures/}}                          % location for color fig.

\newcommand{\etal}{\textit{\mbox{et al.\ }}}                       % et. al.
\newcommand{\trace}{\mbox{Tr}}                                     % Tr
\renewcommand{\Re}{\mathfrak{Re}\,}                                % Re
\newcommand{\captionsize}{\small}

\title{%
\thispagestyle{empty} \vspace{-25mm} 
\begin{flushright}
     \small HU--EP--04/53, LU--ITP 2004/026\\
     \small September 2004
\end{flushright}
\vspace{8mm}
The gluon and ghost propagator and the influence of Gribov copies
\thanks{Talk presented by A.~Sternbeck%
~at the Symposium Lattice 2004, Fermi National Accelerator 
Laboratory, USA%
.}}

\author{A.~Sternbeck\address[HU]{Institut f\"ur Physik, 
        Humboldt-Universit\"at zu Berlin, D-12489 Berlin, Germany},
        E.-M.~Ilgenfritz\addressmark[HU],
        M.~M\"uller-Preussker\addressmark[HU],
        A. Schiller\address[UL]{Universit\"at Leipzig,
        Institut f\"ur Theoretische Physik, D-04109 Leipzig, Germany}}

\begin{document}

\begin{abstract}
  The dependence of the Landau gauge gluon and ghost propagators on the
  choice of Gribov copies is studied in pure $SU(3)$ lattice gauge theory.
  Whereas the influence on the gluon propagator is small, the ghost propagator
  becomes clearly affected by the copies in the infrared region.
  We compare our data with the infrared exponents predicted by the
  Dyson-Schwinger equation approach.\vspace{-0cm}
\end{abstract}

% typeset front matter (including abstract) and HU-EP Number
\maketitle

%-------------------------------------------------------------------------------

The non-perturbative behaviour of the gluon and ghost propagators in
Yang-Mills theories is of interest for the understanding of the mechanism
of confinement of gluons and quarks. In particular the infrared behaviour
of the ghost propagator in the Landau gauge is related to the so-called
Kugo-Ojima confinement criterion~\cite{OjimaKugo-Ojima}, which expresses
the absence of coloured massless asymptotic states from the spectrum of
physical states in terms of the ghost propagator at vanishing momentum.
On the other hand the suppression of the gluon propagator in the infrared
was argued to be related to the gluon confinement~\cite{Gribov}.

Zwanziger~\cite{Zwanziger:2003cf} has suggested that the
behaviour of both propagators in Landau gauge results from the restriction
of the gauge fields to the Gribov region, where the Faddeev-Popov
operator is non-negative. Generically, one gauge orbit has more than
one intersection (Gribov copies) within the Gribov region. In this contribution
we assess the importance of this ambiguity for the ghost and gluon propagator
in $SU(3)$ gauge theory on a finite lattice.

In the continuum the gluon and ghost propagators have been computed from
a coupled and truncated set of Dyson-Schwinger (DS) 
equations~\cite{Alkofer:2000wg}.
Representing the gluon and ghost propagators in the Landau gauge as
\begin{eqnarray}
D_{\mu\nu}^{ab}(q^2)&=&\delta^{ab}
        \left(\delta_{\mu\nu} -{q_{\mu}q_{\nu}/q^2} \right)
        {Z_{gl}(q^2)}/{q^2}, \nonumber \\
G^{ab}(q^2)&=&\delta^{ab} {Z_{gh}(q^2)}/{q^2}
\end{eqnarray}
the low-momentum behaviour of the corresponding dressing functions was
found to be closely constrained by the respective infrared exponents,
$~Z_{gl}\propto(q^2)^{2\kappa},~Z_{gh}\propto(q^2)^{-\kappa}~$
with $~0.5 < \kappa < 1$.

There have been only a few numerical lattice investigations of the $SU(3)$ 
ghost propagator in the past~\cite{Suman:1995zg} contrary 
to the gluon propagator. Here we present data for both propagators
in Landau gauge measured at the same ensemble of $SU(3)$ gauge-fixed 
configurations and concentrate on the infrared behaviour
inside the Gribov region. The strategy of the investigation
is similar to a previous analysis of the $SU(2)$ ghost propagator
\cite{Bakeev:2003rr}.

%------------------------------------------------------------------------------

The Landau gauge condition is implemented by searching for a gauge 
transformation ${}^{g}U_{x,\mu}=g_x U_{x,\mu} 
g^{\dagger}_{x+\hat{\mu}}$ which maximises the gauge functional
$F_{U}[g] \propto  \sum_{x,\mu}\Re \trace \:{}^{g}U_{x,\mu}$
where the gauge field $U_{x,\mu}$ is provided by Monte 
Carlo simulations. This functional has many local extrema whose 
number increases with 
the lattice size and decreasing $\beta \equiv 6/g^2$.
All Gribov copies $\{{}^{g}U\}$ belong to the gauge orbit created 
by $U$ and satisfy the lattice Landau gauge condition
$\partial_{\mu}{}^{g}\!\!A_{x,\mu}=0$ with
\begin{equation}\label{eq:vectorpot}
  {}^{g}\!\!A_{x+\hat{\mu}/2,\mu} = \frac{1}{2i}\left(^{g} U_{x,\mu} - \
                    ^{g} U^{\dagger}_{x,\mu}\right)\Big|_{\rm traceless}.
\end{equation}

Before investigating the infrared behaviour of both propagators we first check
the influence of different ways to select Gribov copies on the average 
propagators taken at some low momenta.

As it was previously found the Gribov copy ambiguity affects the measurements
of the ghost propagator in the case of $SU(2)$ 
\cite{Cucchieri:1997dx,Bakeev:2003rr}. Also from a recent study
the $SU(3)$ gluon propagator is seen to depend systematically 
on the choice of Gribov copies~\cite{Silva:2004bv} contrary
to common belief.
\begin{figure}[t]
  \centering
  \mbox{\hspace{-0.7cm}\includegraphics[width=8.5cm]{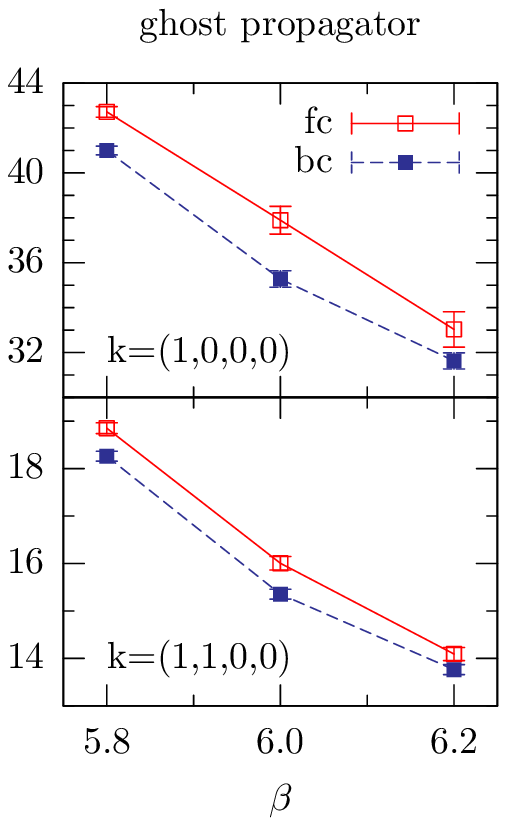}%
        \hspace{-4.9cm}\includegraphics[width=8.5cm]{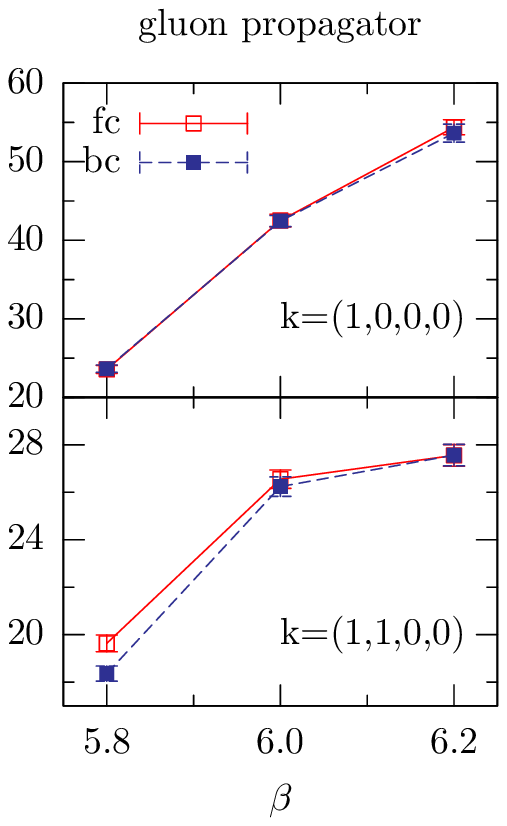}}
  \vspace{-1.5cm}
  \caption{\captionsize 
           The ghost and the gluon propagator $Z(q^2)/q^2$ in lattice units 
           at the two lowest momenta $q(k)$ as a function of $\beta$ for the 
           first (fc) and best (bc) gauge copies (lattice size $24^4$).}
  \label{fig:gh_and_gl_as_func_of_beta}
\vspace{-7mm}
\end{figure}
We report on first measurements on $16^4$ and $24^4$ lattices at $\beta=5.8$, 
6.0, and 6.2 as well as $32^4$ at $\beta=5.8$. In the update the usual 
heatbath algorithm with microcanonical overrelaxation steps is used.
Then for each configuration $N_{cp}=30$ random copies on a $16^4$ lattice
($N_{cp}=40$ on $24^4$ and $N_{cp}=10$ on $32^4$) have been gauge-fixed 
using the standard overrelaxation method until 
$\max_{x}(\partial_{\mu}{}^g\!\!A_{x,\mu})^2 < 10^{-14}$
has been reached.

The propagators were evaluated on the first (fc) and the best (bc) (with respect
to the gauge functional) gauge-fixed copies. The gluon propagator was calculated
for all momenta using the fast Fourier transformation
of the gauge potentials~(\ref{eq:vectorpot}). The ghost propagator 
was obtained using the conjugate gradient method with a plane wave source
only for the lowest non-vanishing momenta.
For each momentum squared,
\mbox{$q^2(k)=(4/{a^2})\sum_{\mu}\sin^2\left({\pi k_{\mu}}/{L_{\mu}}\right)$},
the final propagator is an average over the respective ensemble of gauge-fixed
copies (fc) or (bc) and over different $k$ giving rise to the same $q^2(k)$.
Following Ref.~\cite{Leinweber:1998uu}, only momenta with
$k$ satisfying $\sum_{\mu}k^2_{\mu} - 1/4\cdot(\sum_{\mu}k_{\mu})^2\le 1$
have been used in the analysis of the gluon propagator.

\begin{figure*}[!htb]
  \centering
  \mbox{\hspace{0cm}\includegraphics[width=9.5cm]{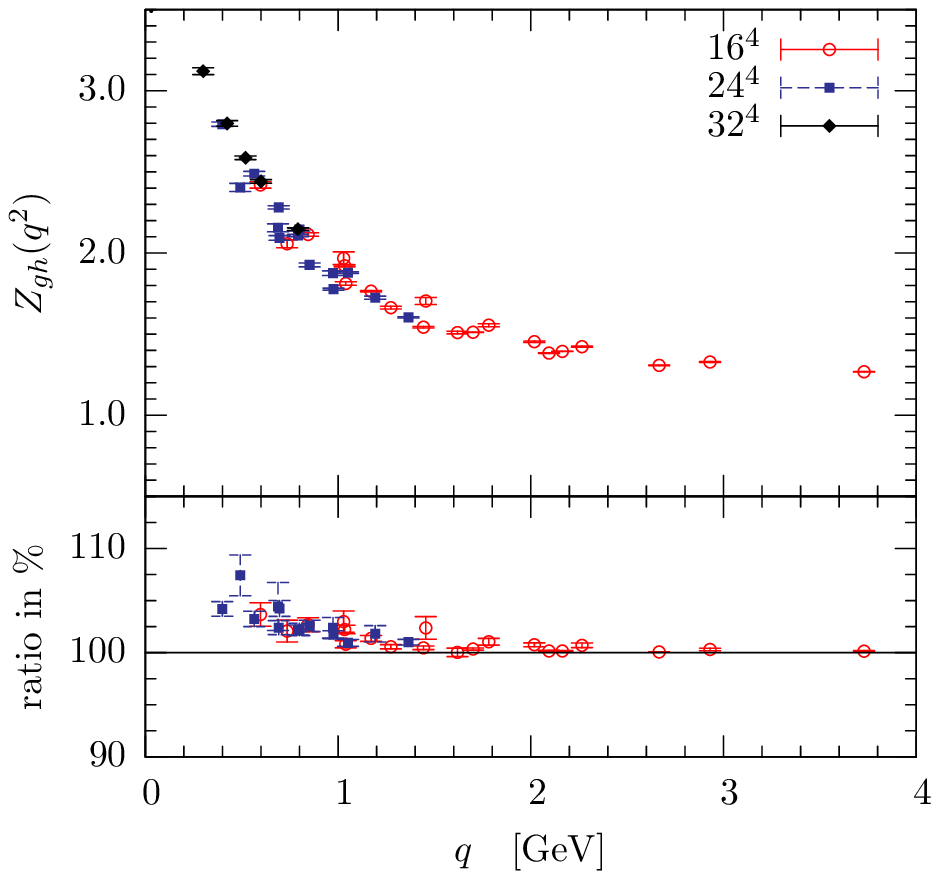}
        \hspace{-2cm}\includegraphics[width=9.5cm]{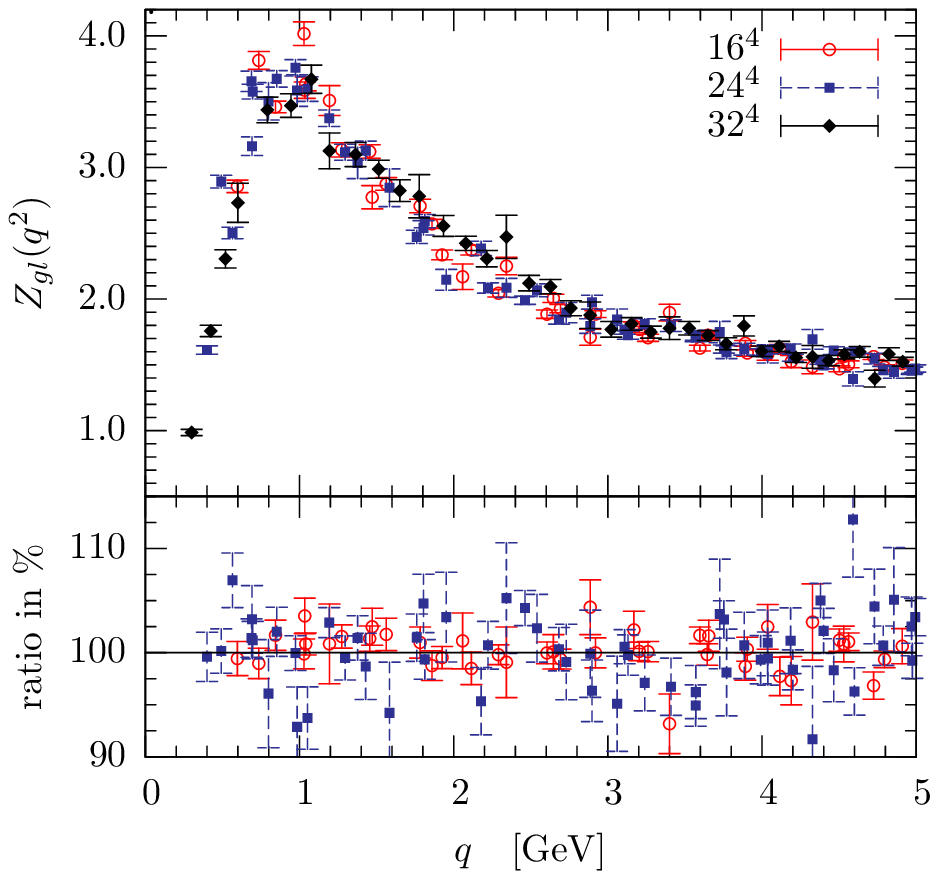}}
  \vspace{-14mm}
  \caption{\captionsize
           The upper figures show the dressing functions of the 
           ghost and gluon propagator measured on the best copies 
           as functions of the momentum $q$
           scaled to physical units for $\beta=5.8, 6.0$ and $6.2$ and 
           various lattice sizes. The lower ones show the ratio 
           $\langle Z^{(\rm fc)}\rangle / \langle Z^{(\rm bc)}\rangle$
           determined from the first (fc) and best (bc) gauge copies.}
     \label{fig:gh_and_gl_as_func_of_q}
\vspace{-3mm}
\end{figure*}

The results from a $24^4$ lattice at the chosen values of $\beta$ are shown
in Fig.~\ref{fig:gh_and_gl_as_func_of_beta} for some lowest momenta.
The ghost propagator is clearly affected by the Gribov
copy problem. The effect is bigger than the statistical error and
increases with decreasing momenta. On the other hand,
for the lowest momentum of the gluon propagator
the impact of Gribov copies is completely below the statistical error.
For other momenta there are some effects visible as 
shown in the lower half of Fig.~\ref{fig:gh_and_gl_as_func_of_q}

%----------------------------------------------------------------------------

The upper half of Fig.~\ref{fig:gh_and_gl_as_func_of_q} shows both propagators
for the best copy as a function of the momentum $q$ scaled with an 
appropriate lattice spacing $a$. In order to compare to other 
studies~\cite{Silva:2004bv,Leinweber:1998uu} we have used $a^{-1}=1.53$, 
1.885 and 2.637 GeV for $\beta=5.8$, 6.0 and 6.2, respectively.
The lower half shows the ratio between the mean value on the first
and best copy of the respective propagators dressing function,
$\langle Z^{(\rm fc)}\rangle/\langle Z^{(\rm bc)}\rangle$. Due to the
low number $N_{cp}$ of gauge copies such ratios on a $32^4$ lattice are not 
shown there. For the lowest momenta the ghost propagator is 
systematically overestimated if measured on an arbitrary (first) 
gauge copy, while for the gluon propagator the statistical noise 
is dominant. However, so far we have only measured the ghost
propagator for a single $k$ without permutations of components.

\begin{figure}[h]
  \vspace{-22mm}
  \hspace{-0.8cm}\includegraphics[width=9.5cm]{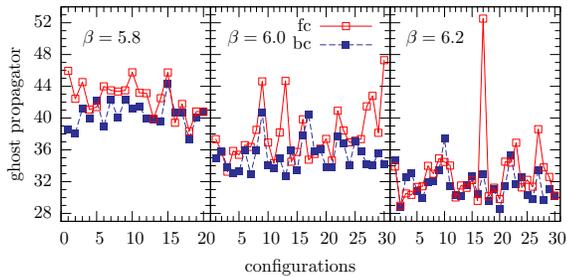}
  \vspace{-15mm}
  \caption{\captionsize 
           The ghost propagator at $k=(1,0,0,0)$ vs.~the first (fc) and
           best (bc) gauge copy on a $24^4$ lattice. }
  \label{fig:ghost_hist}
\vspace{-11mm}
\end{figure}

As in \cite{Bakeev:2003rr} we also found outliers in the data of the ghost 
propagator at the lowest momenta $k=(1,0,0,0)$, shown in 
Fig.~\ref{fig:ghost_hist}, which may make statistical analyses difficult, if
arbitrary (first) gauge copies are studied only. However, the reason is 
unknown yet.

To parametrise the infrared behaviour, we fitted the measured dressing function 
in the form $Z_{gl}=C_{gl}(q^2)^{2\kappa}$ and $Z_{gh}=C_{gh}(q^2)^{-\kappa}$
with common $\kappa$ to the data at some low $q$ values. For the best fit 
with $q_{\rm max}=0.43$ we found $\kappa=0.23(1)$ for the best copy, far 
away from the DS prediction. However, despite $\kappa$ tends to increase
as $q_{\rm max}$ decrease a reasonable statement cannot be made from the 
lattice sizes $L^4$ used.

%-------------------------------------------------------------------------------

In summary, we have reported first results studying the influence 
of the $SU(3)$ ghost and gluon propagators on Gribov copies.
We have found that the ghost propagator is more affected than the gluon 
propagator. Measuring the ghost propagator on an arbitrary first 
gauge-fixed configuration the propagator is systematically overestimated, 
the effect is largest for the lowest momenta. On the other side there is 
an irregular impact on the gluon propagator.
Concerning the infrared behaviour the results from the lattice sizes used
so far do not allow to confirm the proposed scaling behaviour using the
ansatz discussed with an infrared exponent $\kappa>0.5$ simultaneously 
describing both propagators.

%---------------------------------------------------------------------------
\section*{Acknowledgements}

All simulations were done on the IBM pSeries 690 at HLRN. 
We thank H.~St\"uben for contributing parts of the program code 
and V.~Mitrjushkin for discussions.
This work has been supported by DFG under contract FOR 465
(Forschergruppe Gitter-Hadronen-Ph\"{a}no\-meno\-logie).
A.~Sternbeck acknowledges support of the DFG-funded graduate school GK~271. 

%--------------------------------------------------------------------------
% Bibliography:

%=============================================================================

\begin{thebibliography}{1}

\bibitem{OjimaKugo-Ojima}
  T.~Kugo, I.~Ojima, Prog. Theor. Phys. Suppl. {\bf 66}, 1 (1979).

\bibitem{Gribov}
  V.~N.~Gribov, Nucl. Phys. B {\bf 139},1 (1978).

\bibitem{Zwanziger:2003cf}
D.~Zwanziger, Phys. Rev. D {\bf 69}, 016002 (2004) and 
references therein.

\bibitem{Alkofer:2000wg}
R.~Alkofer and L.~von Smekal, Phys. Rept. {\bf 353}, 281 (2001)
and references therein.

\bibitem{Suman:1995zg}
H.~Suman and K.~Schilling, Phys.\ Lett.\ B {\bf 373} 314  (1996);
S.~Furui and H.~Nakajima, Phys.\ Rev.\ D {\bf 69} (2004) 074505.

\bibitem{Bakeev:2003rr}
T.~D. Bakeev, \etal, Phys. Rev. D {\bf 69}, 074507 (2004).

\bibitem{Cucchieri:1997dx}
A.~Cucchieri, Nucl. Phys. B {\bf 508}, 353 (1997).

\bibitem{Silva:2004bv}
P.~J. Silva and O.~Oliveira, Nucl. Phys. B {\bf 690}, 177 (2004).

\bibitem{Leinweber:1998uu}
D.~B. Leinweber, \etal, Phys. Rev. D {\bf 60}, 094507 (1999).

\end{thebibliography}
\end{document}